\documentclass[prl,preprint,superscriptaddress]{revtex4-1}
\usepackage{textcomp}
\usepackage{amssymb,amsmath,MnSymbol,ifsym,latexsym,bm}
\usepackage{wrapfig,subfigure}
\usepackage[Euler]{upgreek}
\usepackage{longtable, supertabular}

\bibpunct[, ]{[}{]}{;}{n}{,}{,}

\usepackage[pdftex]{graphics}
\usepackage[pdftex,final]{graphicx}

\newcommand{\eetap}{\overline{\eta}_\mathrm{p}}

\newcommand{\kB}{k_{\textsc{b}}}
\newcommand{\kBT}{k_{\textsc{b}} \,T}

\newcommand{\edot}{\dot{\varepsilon}}

\newcommand{\Wi}{\mathrm{W\!i}}
\newcommand{\Wics}{\Wi_\mathrm{c-s}}
\newcommand{\Wisc}{\Wi_\mathrm{s-c}}

\newcommand{\NK}{N_\textsc{k}}
\newcommand{\bK}{b_\textsc{k}}

\newcommand{\lamo}{\lambda_\mathrm{0}}

\newcommand{\Mtens}{\bm{\mathrm{M}}}

\newcommand{\bR}{\bm{R}}

\newcommand{\zetac}{\zeta_\mathrm{c}}
\newcommand{\zetaZ}{\zeta_\mathrm{Z}}
\newcommand{\zetaR}{\zeta_\mathrm{R}}

\newcommand{\zetas}{\zeta_\mathrm{s}}

\newcommand{\cdags}{c^\dagger_\mathrm{s}}
\newcommand{\cdagc}{c^\dagger_\mathrm{c}}

\newcommand{\xih}{\xi_\mathrm{h}}
\newcommand{\Nxi}{N_\mathrm{h}}
\begin{document}

\title{Enhancement of coil-stretch hysteresis by self-concentration in polymer solutions}
\author{R. ~Prabhakar}
\affiliation{Department of Mechanical \& Aerospace Engineering, Monash University, Clayton,  VIC 3800, AUSTRALIA\\
email: prabhakar.ranganathan@monash.edu}

\date{\today}

\begin{abstract}
The effect of concentration on coil-stretch hysteresis in extensional flows of polymer solutions is examined with insights from Brownian dynamics simulations of isolated chains and scaling theory for non-dilute solutions. In the hysteresis regime, stretched molecules pervade larger volumes than equilibrium coils. For such chains, intermolecular overlap and hydrodynamic screening crossover set in at concentrations much smaller than the critical overlap concentration $c^\ast$ for equilibrium coils. The width of the hysteresis window is consequently strongly enhanced around $c^\ast$. 
 \end{abstract}

\maketitle

The dynamics of flexible polymers in a solution are determined by the interplay between forces arising from thermal fluctuations, and intramolecular free-energy and hydrodynamic interactions. The latter cause frictional characteristics of molecules to depend on their conformation. The seminal work of \citet{degennes}, \citet{hinch}  and  \citet{tanner} showed that the increase in the average friction coefficient as molecules unravel and stretch leads to a sharp coil-to-stretch transition in extensional flows. The transition occurs at a critical strain rate $\edot$ such that the Weissenberg number $\Wi \equiv \edot \, \lamo$ has a value of $\Wics = 0.5$; $\lamo$ is the characteristic time-scale of  the slowest relaxation mode of a polymer molecule in a quiescent solution.  The coil-stretch transition is also associated with pronounced hysteretic behaviour in conformational and rheo-optical properties  within a window of extension rates such that $\Wisc < \Wi < \Wics$\citep{degennes, hinch, tanner}. The existence of the transition and associated hysteresis in dilute polymer solutions has been demonstrated in experiments and Brownian dynamics simulations with single molecules \citep{perkins, schroeder_science, schroeder} and rheological measurements \citep{sridharPRL}, and is now thought to play an important role in a number of complex flows of dilute polymer solutions \citep{amarouchene, francois}. 

Steady state in extensional flow is primarily the result of a balance between internal resistance of polymer molecules to stretching  and the frictional drag force exerted on molecules by the flowing solvent. Coil-stretch hysteresis in dilute solutions of long flexible molecules stems from the nonlinear dependence of both the entropic resistance and drag forces  on the end-to-end stretch $R$: molecular stiffness begins to diverge as $R$ approaches the contour length $L$, while the friction coefficient changes as the gross molecular shape changes from an isotropic coil at equilibrium to a slender rod when fully stretched.   Within the hysteresis window $\Wisc < \Wi < \Wics$, nonlinearities cause the balance between internal resistance and drag to occur at two distinct values of $R$ \citep{schroeder_science, schroeder, sridharPRL}. One of the stable states corresponds to weakly deformed coils ($R \sim R_0$), and the other is the stretched state ($R \sim L$).  Hysteretic behaviour emerges as large molecules are kinetically trapped in either of  these stable states depending on initial conditions \citep{degennes, schroeder_science, schroeder, sridharPRL, hsiehlarsoncsh}.  The lower bound of the hysteresis window $\Wisc$ can be interpreted as a critical strain rate for an ensemble of stretched molecules to transition to the coiled state. While $\Wics \sim 0.5$ for long molecules, the value of $\Wisc$ depends on molecular weight.  It is known that in a dilute solution, the ratio $\Wics/\Wisc$ is proportional to ratio of the mean friction coefficient of a fully-stretched rod of length $L$,  to the value $\zetaZ$ in the Zimm limit for isolated, isotropic coils at equilibrium of radius $R_0$. Therefore, the size of the hysteresis window, $\Wics/\Wisc \sim \sqrt{\NK}/ \ln \NK$ \citep{schroeder_science, schroeder, sridharPRL}, where $\NK \equiv L^2/R_0^2$ is the number of Kuhn segments in a flexible molecule.

In a highly\textit{ concentrated} polymer solution on the other hand, it is expected that  interpenetration of molecules completely screens out solvent-mediated hydrodynamic interactions \citep{doiedw, rubinsteincolby}. The behaviour of any single molecule thus follows Rouse dynamics as opposed to Zimm hydrodynamics of isolated chains in dilute solutions . Since the Rouse friction coefficient $\zetaR$ of a molecule is independent of chain conformation, the frictional drag force in an extensional flow $F_\mathrm{drag} = \zetaR\,\edot \,R$ is linear with respect to stretch.  No hysteresis occurs in such a case, although the coil-stretch transition is still present at $\Wics = 0.5$ \citep{dpl2}. 

It is therefore natural to ask how coil-stretch hysteresis is affected by concentration as one moves from dilute to concentrated solutions. In polymer solutions close to equilibrium, molecular overlap and interpenetration becomes significant when chain density $c$ exceeds a critical value $c^\ast \sim R_0^{\,-3}$. Intermolecular interactions are typically expected to be negligible in the dilute regime where $c/c^\ast \ll 1$. The objective of this Letter is to firstly show that the phenomenon of coil-stretch hysteresis may have a strong and non-trivial concentration dependence even in solutions nominally considered highly dilute. Secondly,  although hysteresis vanishes as expected in concentrated solutions when  $c/c^\ast \gg 1$, the change in the width of the coil-stretch hysteresis window is non-monotonic with respect  $c/c^\ast$, attaining a large maximum when $c/c^\ast \sim 1$.  This  is relevant to many applications such as turbulent drag reduction, ink-jet printing \emph{etc.} which employ polymeric additives and operate in a range of concentrations where the polymeric solute has little effect on properties under quiescent conditions, while inducing large elastic stresses in flows with significant extensional components. The concentration dependence of viscoelastic phenomena in these applications is far from being fully understood \citep{clasenetal, stoltz}. 

\begin{figure}
\centerline{\resizebox{8.3cm}{!}{\includegraphics{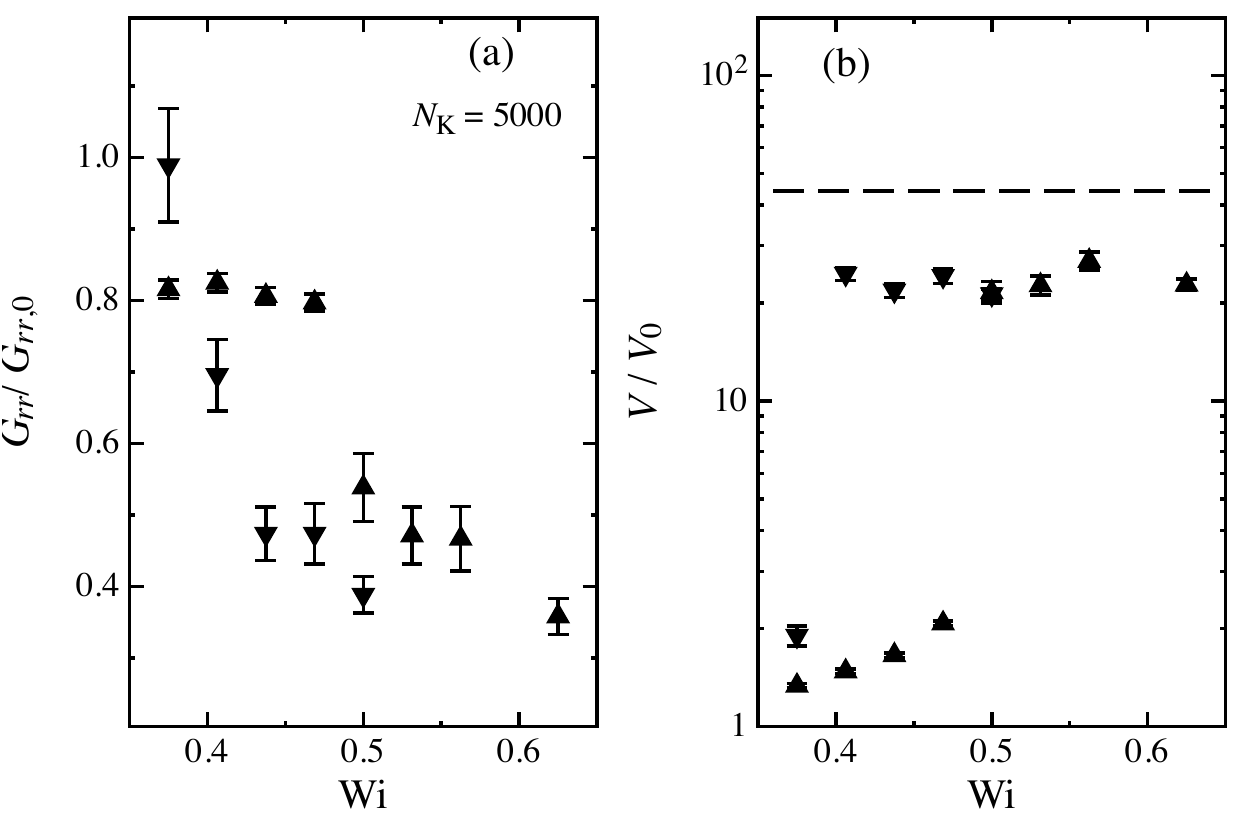}}}
\caption{\label{f:BDSvolfr}  Predictions of Brownian-dynamics simulations \citep{suppmatl} of single-chain dimensions in the vicinity of coil-stretch transition and hysteresis: (a) gyration tensor component $G_{rr}$ of isolated chains transverse to the direction of stretching ($z$) in steady uniaxial extensional flow;  (b) mean pervaded volume in  simulations of isolated linear bead-spring chains. Chain volume is estimated from components of the gyration tensor {$\bm{\mathrm{G}}$} as {$V = G_{rr}\, \sqrt{G_{zz}}$}; {$V_0$} is the equilibrium volume of isotropic coils.  The horizontal dashed line in (b) indicates the volume estimate assuming $G_{zz} = L^2/12$ (valid for rods of length $L$) and $G_{rr} = G_{rr,\,0}/2$. Results are shown for simulations starting with initial ensembles  at equilibrium ($\filledmedtriangleup$), and with initially stretched to $90\%$ maximum permissible length ($\filledmedtriangledown$). These ensembles have different quasi-steady-states within the hysteresis window.} 
\end{figure}

Recent experiments \citep{clasenetal} and  molecular simulations \citep{stoltz} have demonstrated that even in very dilute polymer solutions, flow-induced extension of chains can strengthen the concentration dependence of rheo-optical properties. Insight into the origin of this ability of dilute solutions to ``self-concentrate" can be obtained from observations of average molecular conformation (Fig.~\ref{f:BDSvolfr}) in Brownian-dynamics simulations of isolated chains in extensional flows in the vicinity of the coil-stretch transition \citep{suppmatl}. When $\Wi  \lesssim  1$, conformational fluctuations transverse to the principal stretching axis in uniaxial extensional flow are large and comparable in magnitude to equilibrium fluctuations (Fig.~\ref{f:BDSvolfr} a); this is true of molecules in both the stable coiled and stretched states within the hysteresis window, and at steady states for $\Wi \gtrsim \Wics$.  The ratio $c/c^\ast \sim c \,R_0^3$ can be recognized as the fractional volume pervaded by equilibrium coils. If we set $c^\ast = d_0^{\,-3}$, where $d_0 \equiv R_0/\sqrt{3}$ is the equilibrium size along any fixed direction, the volume fraction pervaded by anisotropic molecules of stretch $R$ and transverse size $d$ is  
\begin{gather}
\phi = c \,R \,d^2 = \frac{c}{c^\ast}\,\frac{R}{d}\,\left(\frac{d}{d_0} \right)^3\,.
\label{e:phi}
\end{gather}
If highly stretched molecules ($R \sim L$) in the hysteresis window are also transversely distended as shown in Fig.~\ref{f:BDSvolfr}  such that the transverse size $d \sim d_0$, Eqn.~\eqref{e:phi} 
suggests that the volume fraction $\phi \gg c/c^\ast$ in a solution consisting of stretched molecules when $\Wi \lesssim 1$.  This supports observations in multi-chain simulations \citep{stoltz} that very dilute solution of equilibrium coils could experience significant intermolecular overlap as chains stretch in the vicinity of the coil-stretch transition. 

To the best of this author's knowledge, a detailed theory for molecular hydrodynamics in a solution of chains of anisotropic conformation  near and beyond overlap has not yet been developed. \citet{clasenetal} suggested that one can nevertheless  take advantage of the concept of ``blobs" in semidilute and concentrated solutions of isotropic coils \citep{degennesbook, doiedw, rubinsteincolby, jain2012, ahlrichs} to understand hydrodynamic screening when dilute solutions self-concentrate in extensional flows. In what follows, a scaling analysis for anisotropic chains is combined with well-known results for  isotropic coils in dilute and non-dilute solutions. While such results are valid well within the corresponding scaling regimes, by its very nature, the problem at hand involves crossovers between isotropic coils and idealized slender rods, and between the dilute and non-dilute regimes. In the absence of detailed simulations or experiments, and for the qualitative exploration intended here,  an interpolation scheme is developed that is consistent with the results of the scaling analysis. An additional simplification here is the neglect of excluded-volume and the effect of $\uptheta$-to-athermal crossover in solvent quality.	

\begin{figure}[h!]
\centerline{\resizebox{8.3cm}{!}{\includegraphics{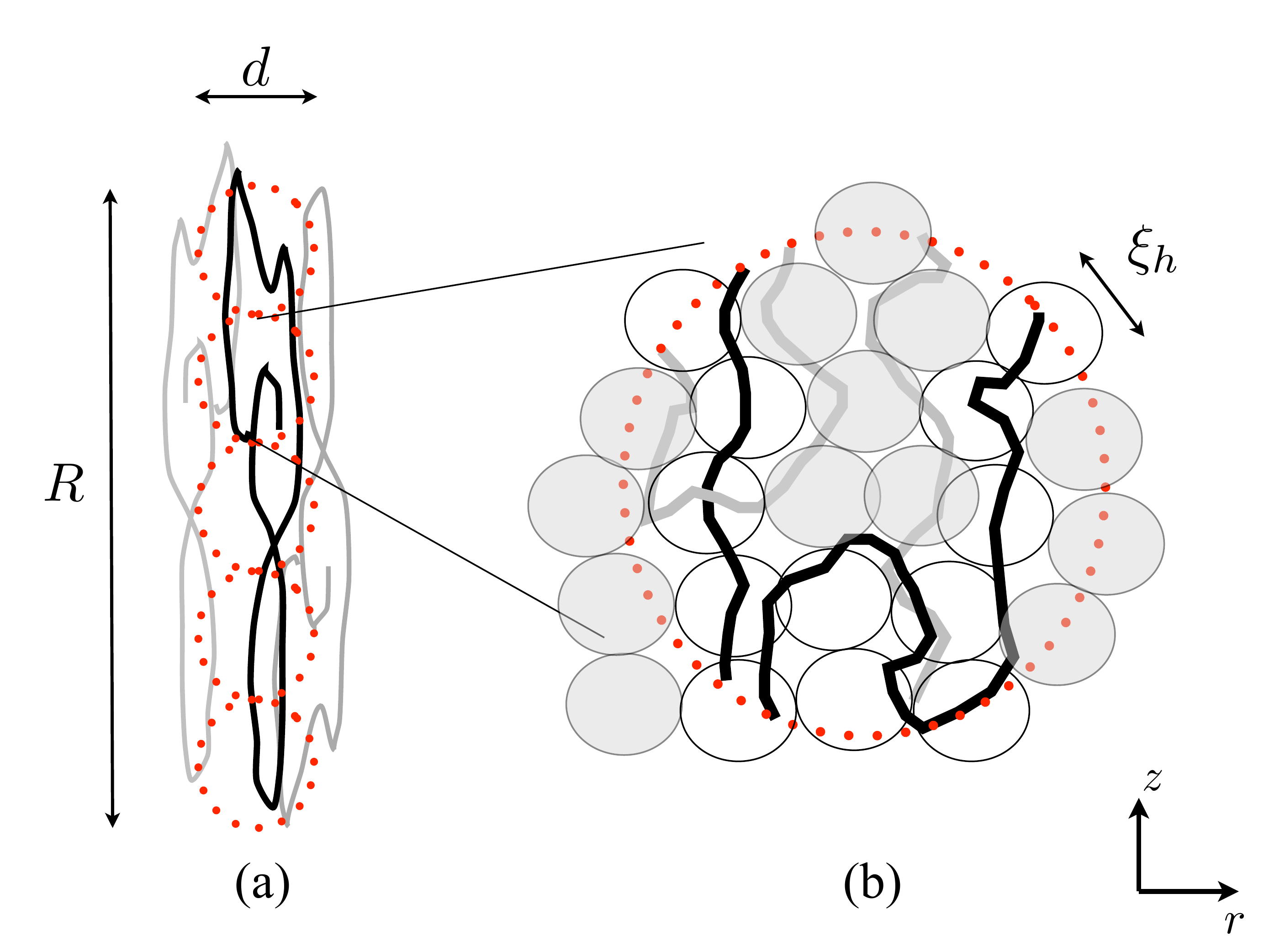}}}
\caption{\label{f:schematic} Schematic of length-scales in a solution of partially unravelled, overlapping polymer chains aligned along the principal stretching direction in a uniaxial extensional flow: each chain of stretch $R$ consists of (a)  $N_d = R/d$ domains of size $d$ determined by transverse fluctuations, and  (b) $\Nxi$ correlation blobs of size $\xih$. Correlation blobs from all chains fill space exactly.}
\end{figure}

In a solution of anisotropic, partially-stretched chains of aspect ratio $R/d$ oriented along the extensional axis, intermolecular screening of hydrodynamic interactions on molecular overlap is described in terms of a screening length $\xih < d$ such that intramolecular hydrodynamic interactions persist only at  length scales smaller than $\xih$,  but are screened at larger length scales. The quantities $\xih$ and $\Nxi$ are determined by firstly noting that $\xih$ is the scale at which segments in any chain first significantly encounter those from  neighbouring chains; hence $c \,\xih^3 \Nxi = 1$. Secondly, it is assumed that the equilibrium ideal random-walk structure persists at length scales smaller than $d$, and $\xih^2/d^2 = (\NK/ \Nxi)\,/\,(\NK/ N_d) = N_d/\Nxi$ where  $N_d = R/ d$ (Fig.~\ref{f:schematic}). When chains overlap therefore, $\xih/d = \phi^{-1} $ and $\Nxi = \phi^2 \,(R/ d)$. When the pervaded volume fraction (Eq.~\eqref{e:phi}) $\phi \geq 1$, molecules on average act as freely-draining Rouse chains of $\Nxi$ ``correlation blobs" (Fig.~\ref{f:schematic}; \citep{degennesbook,doiedw}), each blob with a friction coefficient $(\xih/d_0)\,\zetaZ$.  In such a case, the overall Rouse-like friction coefficient $\zetas$ of a partially-stretched chain is such that 
\begin{gather}
\frac{\zetas}{\zetaZ} = \frac{\xih}{d_0}\, \Nxi  = \phi\, \frac{R}{d_0}  = \frac{c}{c^\ast}\, \left(\frac{R}{d_0}\right)^2 \,  \left(\frac{d}{d_0}\right)^2\,.
\label{e:semidilute}
\end{gather}
For\textit{ isotropic} coils, on the other hand, the pervaded volume fraction  is $c/c^\ast$.  When $c/c^\ast < 1$, the  friction coefficient for coils is only weakly dependent on concentration, and $\zetac \sim \zetaZ$ \citep{doiedw, clasenetal, liuetal}.  As in the case of stretched chains,  hydrodynamic screening leads to a Zimm-to-Rouse crossover when coils overlap; substituting $R = d = d_0$ in Eq.~\eqref{e:semidilute} gives
\begin{gather}
\frac{\zetac}{\zetaZ} = \frac{c}{c^\ast} \,, \quad \text{when } c/c^\ast > 1 \,.
\label{e:semidilute2}
\end{gather}

In either coiled or stretched chains, hydrodynamic screening is complete when $\zetas$ or $\zetac$ attain the Rouse value $\zetaR$, which is the bare friction of a chain of $\NK$ Kuhn segments, each of length $\bK \equiv R_0^2/L$. Therefore,  both  $\zetas/\zetaZ$ and $\zetac/\zetaZ$ have an upper bound of $\zetaR/\zetaZ = (\bK/d_0)\,\NK = \sqrt{3\,\NK}$. Using this estimate in Eq.~\eqref{e:semidilute2} above, the Zimm-to-Rouse crossover is complete for isotropic coils at a concentration $\cdagc$ such that $\cdagc/c^\ast = \zetaR/\zetaZ = \sqrt{3 \NK}$. For stretched molecules, Eq.~\eqref{e:semidilute} suggests that $\zetas = \zetaR$ at a different concentration $\cdags$ such that
\begin{gather}
\frac{\cdags}{c^\ast} \,=  \,\left(\frac{R}{d_0}\right)^{-2} \,  \left(\frac{d}{d_0}\right)^{-2} \,\frac{\cdagc}{c^\ast}  \,.
\end{gather}
For transversely-distended chains in the stretched state in coil-stretch hysteresis, not only is the crossover for partially stretched chains complete at $ \cdags \ll \cdagc$, but it occurs well within the dilute regime at $\cdags/c^\ast \sim \NK^{-1/2} \ll 1$. 

The scaling analysis above suggests that there exists a large range $\NK^{-1/2} < c/c^\ast < 1$ in which the solution is dilute in the coiled state and $\zetac = \zetaZ$, but is concentrated with complete screening of hydrodynamic interactions in the stretched state, and $\zetas = \zetaR$.  It is only when  $c/c^\ast  > 1$ that $\zetac$  begins to ``catch up" with $\zetas$. The frictional drag ratio $\zetas/\zetac$ in the self-concentrated regime thus scales as $\zetaR/\zetaZ \sim \sqrt{\NK}$. In comparison, the stretched-to-coiled friction ratio in the limit of infinite dilution scales as $\sqrt{\NK}/\ln \NK$ as pointed out earlier. Thus, in moving from very dilute to highly concentrated solutions, the friction coefficient ratio first increases from $\sqrt{\NK}/\ln \NK$ to a maximum of $\sqrt{\NK}$ in the self-concentrated regime, and then decreases to unity. It is expected that the width of the coil-stretch hysteresis window will follow suit. The increase in hysteresis width by a factor proportional to $\ln \NK$ can be quite significant in flexible polymers with $\NK \gtrsim 10^3$ \citep{clasenetal}.  

\begin{figure*}
\centerline{\resizebox{16.6cm}{!}{\includegraphics{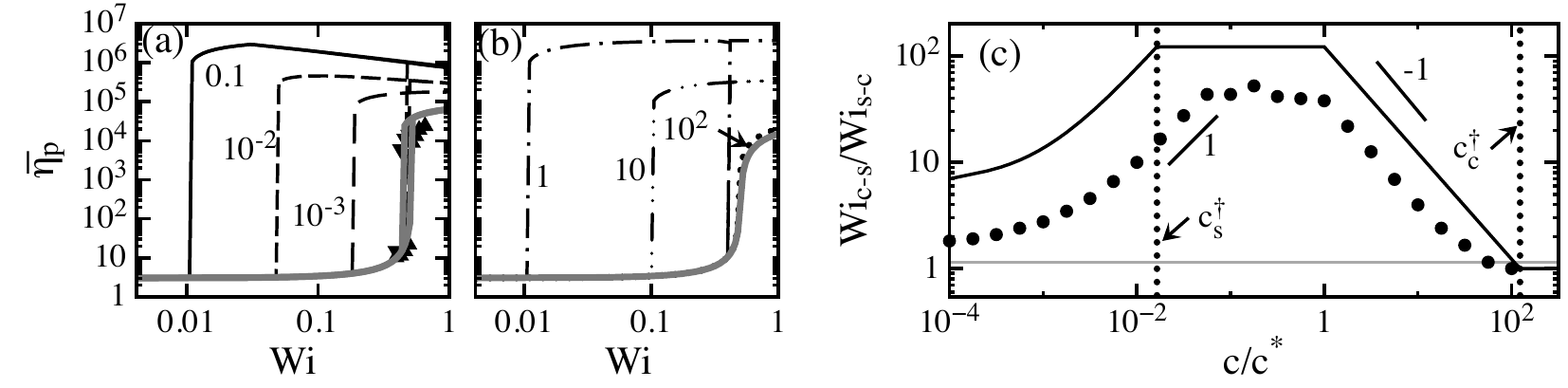}}}
\caption{\label{f:csh} Predictions of the conformation-tensor model for coil-stretch hysteresis in the polymer contribution to steady-state extensional viscosity (scaled by $c\, \kBT \lamo$) for $\NK = 5 \times 10^3$,  hown for the sake of clarity separately for (a) $c/c^\ast < 1$, and (b)  $c/c^\ast \geq 1$; numbers alongside curves indicate $c/c^\ast$ values. Symbols shown in (a) are data for the window predicted by Brownian dynamics simulations \cite{suppmatl}. Thick gray curves in (a) and (b) are, respectively, predictions obtained with the model in the infinite-dilution limit, and that obtained by the FENE-P model \citep{wedgewood} with conformation-\emph{independent} friction. }
\end{figure*}

Going beyond scaling analysis to obtain numerical predictions for hysteresis in macroscopic properties such as the extensional viscosity requires two additional modeling constructs. Firstly, the average friction coefficient $\zeta$ has to be quantitatively related to molecular conformation and concentration in a manner consistent with the scaling results presented above. Secondly, prediction of hysteresis requires this model for friction to be considered together with the non-linear stretch-dependence of entropic resistance.  To achieve the first objective for a solution of partially-unravelled chains of mean stretch $R$ and transverse size $d$, a linear ``mixing rule" is used here to interpolate $\zeta$  between an estimate $\zetas$ for penetrable rods of length $R$ and diameter $d$, and  $\zetac$, the estimate for isotropic coils of size $d_0$   \citep{suppmatl}. The estimate $\zetas$ in turn uses a second interpolation  \citep{suppmatl} to connect  expected concentration dependence for a dilute suspension of rods obtained from Batchelor's theory \citep{batchelor} when $\phi \ll 1$, and for overlapping anisotropic chains as described above (Eqn.~\eqref{e:semidilute}). A single unknown prefactor in this model is determined by requiring predictions of hysteresis with this model at infinite dilution match those obtained with Brownian dynamics simulations of isolated bead-spring chains \citep{suppmatl}. 

To obtain hysteresis predictions, a Fokker-Planck equation can be solved for the probability distribution of the end-to-end vector $\bR$ of a single test chain in the polymer solution  \citep{degennes, schroeder}. An alternative but related approach is to invoke Peterlin-type \citep{dpl2, birdwiest, wedgewood} closure approximations to solve a set of coupled ordinary differential equations for the second-moment or the so-called conformation tensor $\Mtens \equiv \langle \bR \bR \rangle$, the angular brackets denoting an ensemble average. This approach has been demonstrated to yield comparable results and has been widely used to understand coil-stretch hysteresis \citep{degennes, hinch, tanner, phanthien, dunlapleal}, and is followed here. The equations for the evolution of the axial ($zz$) and transverse ($rr$) components of the conformation tensor  $\Mtens$ are \citep{dpl2}:
\begin{eqnarray}
\frac{d M_{zz}}{d t} = 2 \,\edot\, M_{zz} - \frac{4 H}{\zeta} \, M_{zz} + \frac{4 \kBT}{\zeta} \,,\\\label{e:mxxode}
\frac{d M_{rr}}{d t} = - \edot\, M_{rr} - \frac{4 H}{\zeta} \, M_{rr} + \frac{4 \kBT}{\zeta} \,.
\label{e:myyode}
\end{eqnarray}
The nonlinear effective stiffness of resistance to stretching,  $H = H_0 \,(L^2 - R_0^2) \, [L^2 - (M_{zz} + 2 M_{rr})]^{-1}$, where $H_0 = 3 \,\kBT/ R_0^2$, $\kB$ is the Boltzmann constant, and $T$ is the absolute solution temperature. The friction coefficient $\zeta$ is calculated as described above \citep{suppmatl} after identifying the instantaneous $R$ and $d$ of chains with $M_{zz}^{1/2}$ and $M_{rr}^{1/2}$, respectively.  The near-equilibrium relaxation time $\lamo = \zetac\, R_0^2\,/ \,(12\, \kBT)$, and varies  with concentration as $\zetac$.  On rescaling model equations with $R_0$ and $\lamo$ as the characteristic length and time scales, the sole dimensionless model parameters are $\NK$, $c/c^\ast$ and $\Wi$. The dimensionless polymer contribution to the extensional viscosity, $\eetap = H (M_{zz} - M_{rr})/(\kBT \, \Wi)$ \citep{dpl2}. Hysteresis predictions are obtained by integrating to steady-state at each value of $\Wi$, starting with equilibrium, and with highly-stretched initial conditions. 

The variation of hysteresis windows with concentration predicted thus in Fig.~\ref{f:csh} (a) and (b) for $\NK = 5 \times 10^3$ is consistent with the expectation from scaling analysis that the hysteresis window widens as the stretch-to-coil transition $\Wisc$ decreases with increasing concentration in the dilute regime (Fig.~\ref{f:csh} (a)) until about $\cdags/c^\ast  \sim 0.1$. At concentrations above $c/c^\ast = 1$ (Fig.~\ref{f:csh} (b)), $\Wisc$ begins to increase towards $\Wics = 0.5$ and hysteresis becomes weaker, vanishing completely at $c/c^\ast \sim 100$. The non-monotonic concentration dependence of $\Wics/\Wisc$ predicted by the conformation-tensor model in Fig.~\ref{f:csh} (c) compares well with an upper bound estimated from the ratio $\zetas/\zetac$ calculated as a function of concentration after assuming $R = R_0$ (equilibrium dimensions) for the coiled state, and $R = L$ (fully-stretched) and  $d^2 = d_0^2/2$ (large transverse size) for the stretched state. Similar arguments are expected to hold when excluded-volume interactions are important. Indeed, the self-concentration effect may be marginally greater because the hysteresis window size at infinite dilution (horizontal line in Fig.~\ref{f:csh} (c) ) is smaller in good solvents and may even vanish completely  \citep{shika, radhaunderhill}, but $\zetaR$ (and hence the maximum in Fig.~\ref{f:csh} (c) ) is expected to be unaffected by solvent quality. 
 

The results presented above are consequences of the hypothesis that hydrodynamic screening concepts, usually invoked to analyze semi-dilute solutions near equilibrium, can also be used to understand their non-equilibrium behaviour. This is of fundamental value: validation of the results here against experiments and simulations would lay the foundation for a more comprehensive understanding of polymer solutions; on the other hand, if experiments and simulations do not agree with the results here, it would show what is lacking in our knowledge of screening and the Zimm-to-Rouse crossover.  There appears to be experimental evidence for the predictions above from measurements of the characteristic time-scale of radial decay in exponentially thinning slender filaments of dilute polymer solutions. \citet{clasenetal} have observed this time-scale to be strongly enhanced with concentration even for solutions with $c/c^\ast \ll 1$, while simulations show it to be proportional to the hysteresis window size $\Wics/ \Wisc$ \citep{prabhakar}. Firm confirmation may be obtained by more systematic rheological measurements  \citep{sridharPRL} and by visualizing stained molecules in extensional flows \citep{schroeder_science}. The predictions here may further be tested in large-scale simulations, when it becomes feasible to handle multiple, hydrodynamically-interacting chains long enough to resolve all relevant length and time scales in elongational flows with appropriate periodic boundary conditions \citep{stoltz}.

\acknowledgments{This work was supported by a CPU-time grant on the National Computational Infrastructure at the Australian National University, Canberra.}
\bibliography{bib}
\bibliographystyle{aipnum4-1}
\end{document}


\title{Supplementary material for ``Enhancement of coil-stretch hysteresis by self-concentration in polymer solutions"}

\author{R. ~Prabhakar}
\email{prabhakar.ranganathan@monash.edu}
\affiliation{Department of Mechanical \& Aerospace Engineering, Monash University, Clayton,  AUSTRALIA}

\date{\today}

\pacs{}
\maketitle

\section{\label{s:zeta} Conformation- and concentration-dependent friction coefficient}

The average friction coefficient $\zeta$ of partially unravelled and partially aligned chains in a solution is a function of average chain dimensions --- the stretch $R$ and transverse size $d$ --- and the chain number density $c$.  The stretch $R$ can vary between $d_0$  at equilibrium and  the polymer contour length, $L$. The range for the transverse chain size $d$ is in principle $\bK < d < d_0$. However, in extensional flows where $\Wi \lesssim 1$, $d  \lesssim d_0$. The Kuhn-segment length, $\bK \equiv R_0^2/L$, and the number of Kuhn segments in a chain, The $\NK \equiv L^2/R_0^2$.

The function $\zeta$ is anticipated to lie between the friction $\zetac$ for isotropic coils of size $d_0 \equiv  R_0/\sqrt{3}$  at the same concentration at equilibrium, and an estimate  $\zetas$ derived assuming that chains are slender rod-like objects  of average length $R$ and $d$, all aligned in the principal stretching direction in a flow.  A simple linear ``mixing rule" interpolates $\zeta$ between $\zetac$ and $\zetas$ with respect to $R$ as follows: 
\begin{gather}
\frac{\zeta}{\zetaZ}  = \left(\frac{\zetas}{\zetaZ} \right)\, \frac{ R - d_0}{L - d_0} \,+\, \left(\frac{\zetac}{\zetaZ}\right) \,\frac{ L - R}{L - d_0} \,,
\label{e:zetamixrule}
\end{gather} 
where $\zetaZ$ is the limiting (``Zimm") value for \textit{isolated} coils at equilibrium. The two ratios $\zetac/\zetaZ$ and $\zetas/\zetaZ$ in the equation above  are themselves concentration dependent, and the latter is also conformation dependent.  These relationships are modeled by interpolating between scaling results expected under different conditions.

\subsection{Concentration dependence of $\zetac/\zetaZ$}
The variation of  the friction coefficient $\zetac$ of equilibrium coils due to intermolecular hydrodynamic interactions in the dilute regime when $c < c^\ast$ is not exactly known, but from the observed concentration dependence of the largest relaxation time $\lamo \sim \zetac\, R_0^2/\, \kBT$ in small-amplitude-oscillatory-shear experiments \citep{clasenetal, liuetal}, $\zetac$ is expected to vary only weakly with concentration until $c^\ast$ is approached. Beyond $c^\ast$, the effects of hydrodynamical screening and crossover  in dynamical properties in the linear response regime (\emph{e.g} diffusivity) have been well documented \citep{rubinsteincolby}. These observations suggest that the variation in $\zetac$ with concentration may be approximated as follows:
\begin{align}
\frac{\zetac}{\zetaZ} = \begin{cases} 1 \,, \quad \text{if } c \leq c^\ast\,, \\[2mm]
\displaystyle{\frac{c}{c^\ast}} \,, \quad \text{if } c > c^\ast\,,
\end{cases}
\end{align} 
the second result following from applying blob arguments for equilibrium coils, as highlighted in the main text. Hydrodynamic screening is considered to be complete when $\zetac$ reaches an upper limit set by the Rouse friction coefficient, which for a chain of $\NK$ Kuhn-segments, is  
\begin{gather}
\frac{\zetaR}{\zetaZ} = \frac{\bK}{d_0}\,\NK = \sqrt{3\,\NK}\,
\end{gather} 

\subsection{Concentration and conformation dependence of $\zetas/\zetaZ$}

Three distinct concentration regimes can be identified for modeling $\zetas/\zetaZ$ depending on the instantaneous pervaded volume fraction of stretched molecules:
\begin{gather}
\phi = c \,R \,d^2 = \frac{c}{c^\ast}\,\frac{R}{d}\,\left(\frac{d}{d_0} \right)^3\,.
\label{e:phidef}
\end{gather}
where the overlap concentration is estimated as  $c^\ast = d_0^{\,-3}$.

\subsubsection{$\zetas/\zetaZ$ in dilute regime, $\phi \leq 10^{-2}$}
In a suspension of aligned, non-overlapping rods, the mean transverse separation $h$ is such that $c \,R\, h^2 = 1$, and hence from Eqn.~\eqref{e:phidef},
\begin{equation}
\frac{h}{d} = \frac{1}{\sqrt{\phi}} \,.
\label{e:heqn}
\end{equation}
The following approximation is based on Batchelor's \citep{batchelor} results for  suspensions of slender rods:
\begin{equation}
\frac{\zetas}{\zetaZ}  = \frac{K}{K +  \,\ln F} \, \left(\frac{R}{d_0}\right)\,,
\label{e:zetadilutebatch}
\end{equation}
where the argument of the logarithmic correction in the denominator is
\begin{align}
F \,=\, \begin{cases}
\,\displaystyle{\frac{R/d}{1 + R/h} \, = \, \frac{R/d}{1 + (R/d) \sqrt{\phi}}}\,, &\text{if } R/d \,>\,(1 - \sqrt{\phi})^{-1}\\[5mm]
\,1 \,.& \text{if } R/d \,\leq \,(1 - \sqrt{\phi})^{-1}\\
\end{cases}
\label{e:fcases}
\end{align}

The expressions above recover expected behaviour: $\zetas \rightarrow \zetaZ$ for isotropic coils when $R \rightarrow d_0$, and for high aspect ratios when $\ln F \gg K$,
\begin{gather}
\frac{\zetas}{\zetaZ} = \frac{K}{\ln (R/d) - \ln (1 + R/h)} \, \frac{R}{d_0}\,,
\label{e:batchelor}
\end{gather}
as suggested by Batchelor's results, which are in turn expected to be valid when $h/d \gg 1$ or $\sqrt{\phi} \ll 1$ \citep{batchelor}. Therefore, Eqns.~\eqref{e:zetadilutebatch} and \eqref{e:fcases} are used when $\sqrt{\phi} \leq 0.1$. The conditional specification of $F$ in Eqn.~\eqref{e:fcases}  avoids singular behaviour as the aspect ratio $R/d$  unity in the limit of infinite dilution,  $\phi \rightarrow 0$; the choice of $F =1$ is only needed for $R/d \approx 1$ which typically occurs near equilibrium, where the  mixing rule in Eqn.~\eqref{e:zetamixrule} ensures that $\zetas$ is of little consequence. The unknown prefactor $K$ is fixed at a value of 0.15 so that the hysteresis loop in intrinsic extensional viscosity predicted by the model here when $c = 0$  is comparable in size to that obtained in Brownian dynamics simulations for isolated bead spring chains (see Fig.~\ref{f:BDScsh} below). 

\subsubsection{$\zetas/\zetaZ$ in non-dilute regime, $\phi \geq 1$}
\begin{figure}[h!]
\centerline{\resizebox{12.7cm}{!}{\includegraphics{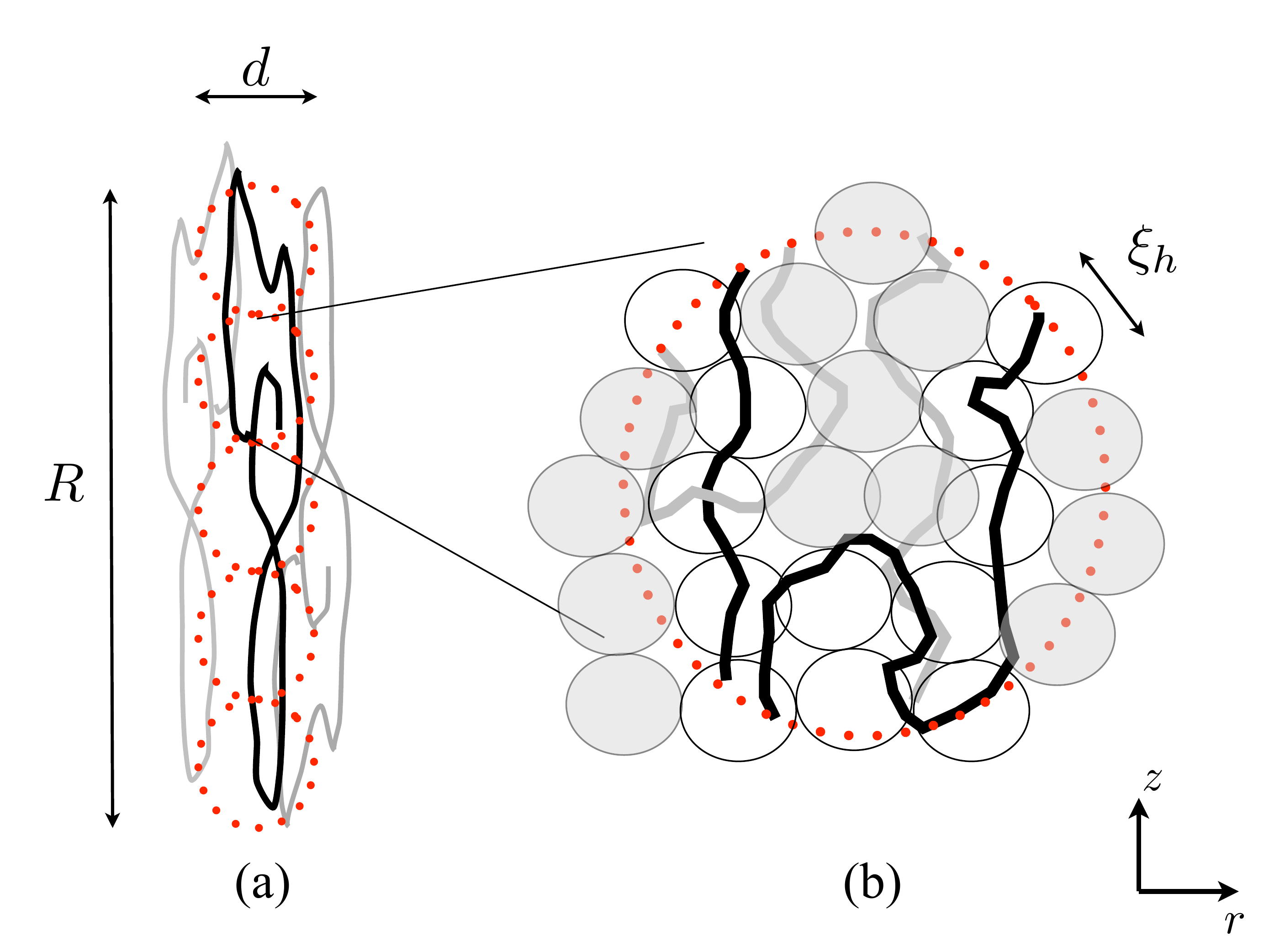}}}
\caption{\label{f:schematic} Schematic of length-scales in a solution of partially unravelled, overlapping polymer chains aligned along the principal stretching direction in a uniaxial extensional flow }
\end{figure}
Each molecule is a chain of $\Nd = R/d$ ``beads". Within each bead are ``correlation  blobs" \citep{degennesbook, rubinsteincolby} of size $\xi$, with a total of $\Nxi$  blobs along the chain. Intramolecular hydrodynamic interactions are restricted to segments within a blob, but are screened at larger length scales. The hydrodynamic screening length $\xih$ is that at which the intramolecular segmental density is equal to the average segmental density of the solution as a whole, which in turn implies that blobs of all molecules exactly fill space, or
\begin{equation}
c \,\Nxi\, \xih^3 \,=\, 1\,.
\label{e:blob1}
\end{equation}
Assuming that an equilibrium-like structure survives at length scales smaller than the size of the transverse fluctuations $d$ when $\Wi \lesssim \mathrm{O}(1)$,  for solutions near the $\Theta$-state,
\begin{equation}
\frac{\xih^2}{d^2} =  \, \frac{\NK/\Nxi }{\NK/\Nd} \, =  \, \frac{R/ d}{\Nxi}  \,.
\label{e:blob2}
\end{equation}
The  rod-like mean friction coefficient of the partially-stretched chain chain is calculated as the Rouse drag of $\Nxi$ blobs, each with a Zimm friction $\zetaxi = (\xih/d_0) \,\zetaZ$, 
\begin{equation}
\frac{\zetas}{\zetaZ} = \frac{\xih}{d_0} \Nxi = \phi \, \frac{R}{d_0} \,,
\label{e:zetasemidil}
\end{equation}

\subsubsection{$\zetas/\zetaZ$ in transition regime, $0.01 < \phi < 1$}
In this regime, another interpolation is used for $\zetas/\zetaZ$ between the value predicted at  $\phi = 1$
by  Eqn.~\eqref{e:zetasemidil}:
\begin{equation}
\frac{\zetas}{\zetaZ} = \Nd\, \frac{d}{d_0} = \frac{R}{d_0} \,, 
\label{e:zetaovlp}
\end{equation}
and the value calculated by Eqn.~\eqref{e:zetadilutebatch} when $\phi = 0.01$:
\begin{equation}
\frac{\zetas}{\zetaZ} =  \frac{K}{K + \,\ln \left.F\right|_{\phi = 0.01}} \,,
\end{equation}
where $\left.F\right|_{\phi = 0.01}$ is calculated from Eqn.~\eqref{e:fcases} for any given $R$ and $d$, but with $\phi = 0.01$. Hence, given $c/c^\ast$, and the dimensions $R$ and $d$, if $0.01 < \phi < 1$, a linear interpolation between the two values of $\zetas/\zetaZ$ above is:
\begin{equation}
\frac{\zetas}{\zetaZ} = \left[ \frac{K}{K + \,\ln \left.F\right|_{\phi = 0.01}} \, \frac{1 - \phi}{1 - 0.01} \,+\, \frac{\phi - 0.01}{1 - 0.01} \right] \, \frac{R}{\,d_0} \,.
\label{e:zetatrans}
\end{equation}

\section{\label{s:bds} Brownian dynamics simulations of isolated flexible polymers}

\begin{figure}
\centerline{\resizebox{8.3cm}{!}{\includegraphics{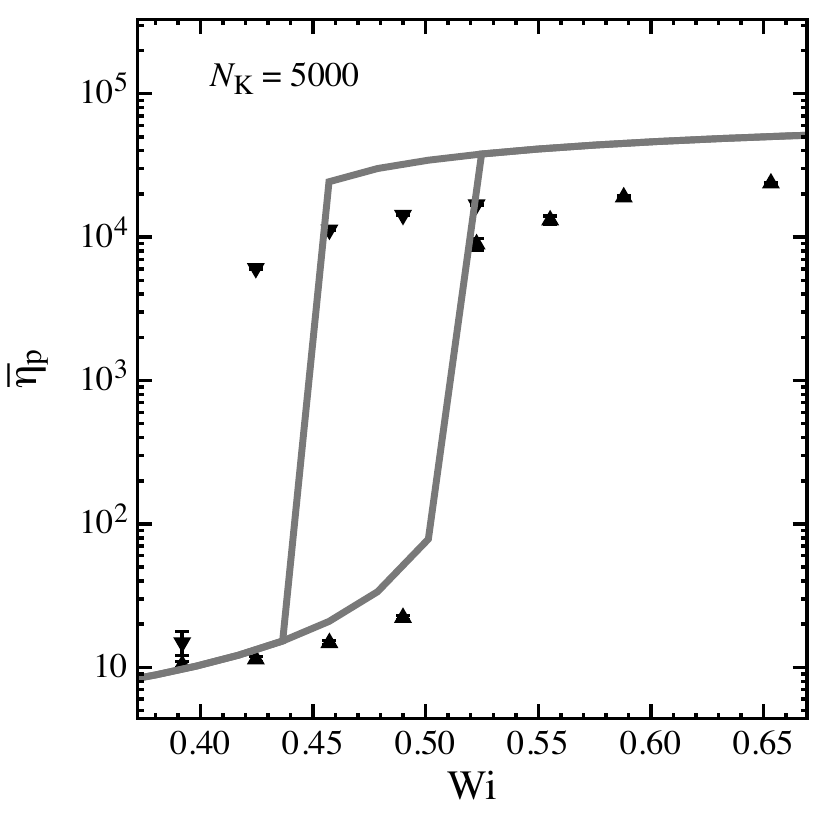}}}
\caption{\label{f:BDScsh}  Coil-stretch hysteresis in steady-state extensional viscosity: symbols are results of Brownian dynamics simulations with isolated 26-bead chains, obtained with initial ensembles consisting of equilibrium coils ($\filledmedtriangleup$), and consisting of initially stretched chains ($\filledmedtriangledown$); continuous curves are predictions of the conformation-tensor model at infinite dilution.} 
\end{figure}

Results of single-molecule Brownian dynamics simulations presented in this study are from Ref.~\citep{thesis}. The molecular model and its parameters, and the simulation algorithm have been reported in Refs.~\citep{prabhakarmultiplicative, prabhakarSFG}. Briefly, polymer molecules of $\NK$ Kuhn segments are modeled as chains of $N < \NK$ beads connected by Finitely-Extensible Nonlinear Elastic (FENE) springs, and intramolecular hydrodynamic interactions are incorporated through Rotne-Prager-Yamakawa tensors. Excluded-volume interactions are neglected for chains at the $\Theta$-state. Independent trajectories of chains are generated by integrating a stochastic differential equations for bead positions that describe their motion under the combined action of hydrodynamic drag forces exerted by a homogeneous flow of the surrounding solvent and its thermal fluctuations, and intramolecular forces arising from free-energy interactions. The gyration tensor 
\begin{gather}
\bm{\mathrm{G}} \equiv \frac{1}{N} \sum_{\nu = 1}^{N} \langle \bm{R}_\nu
\bm{R}_\nu \rangle\,,
\end{gather}
where $\bm{R}_\nu$ is the position vector of the $\nu$-th bead relative to the center-of-mass of a chain of $N$ beads, and the angular brackets represent averaging over a large ensemble of independent stochastic realizations of chain configurations. The volume pervaded by a chain $V \equiv \sqrt{\det{\bm{\mathrm{G}}}}$. In flow, the polymer contribution to the extra-stress $\bm{\uptau}_\mathrm{p}$ is given by the Kramers' equation \citep{dpl2}, and the dimensionless extensional viscosity is defined as
\begin{align}
\eetap\, & \equiv\, -\, \frac{\tau_{\mathrm{p},\,zz}-
\tau_{\mathrm{p},\,rr}}{c\, \kBT \,\Wi}\,. \label{e:eeta}
\end{align}
The evaluation of the longest relaxation time for defining the Weissenberg number $\Wi$ is described in Ref. \citep{prabhakarSFG}. Predictions of coil-stretch hysteresis are obtained by running to steady-state two separate ensembles with different initial distributions of chain configurations at each $\Wi$: results for coiled states are obtained with an initial ensemble at  equilibrium, while stretched state simulations start with chains nearly fully extended and all oriented in the $z$-direction. Figure~\ref{f:BDScsh}  shows hysteresis in $\eetap$ obtained with simulations, along with predictions of the conformation-tensor model in the limit of infinite dilution, with $K = 0.15$.

\bibliography{bib}
\bibliographystyle{aipnum4-1}